\documentclass[conference,letters]{IEEEtran}
\IEEEoverridecommandlockouts

\usepackage[hidelinks]{hyperref}
\usepackage[cmex10]{amsmath}
\usepackage{amssymb,amsfonts}
\usepackage{dblfloatfix}

\usepackage[ruled,vlined]{algorithm2e}
\usepackage{graphicx}
\usepackage{float}
\graphicspath{{Figures/PDF/}{Figures/PNG/}}

\usepackage{booktabs}
\usepackage{siunitx}
\usepackage[numbers,compress]{natbib}
\usepackage{texnames}
\usepackage{multirow, bigstrut, makecell}
\usepackage{bm,bbm}
\usepackage{orcidlink}

\makeatletter
\newcommand{\linebreakand}{%
  \end{@IEEEauthorhalign}
  \hfill\mbox{}\par
  \mbox{}\hfill\begin{@IEEEauthorhalign}
}
\makeatother

\begin{document}

\title{\uppercase{Single-Step Latent Consistency Model for Remote Sensing Image Super-Resolution}
\thanks{*Corresponding author: Hanlin Wu. This work was supported in part by the National Natural Science Foundation of China under Grant 62401064 and Grant 62101052, and in part by the Fundamental Research Funds for the Central Universities under Grant 2024TD001 and Grant 2024JX069.}
}

\author{\IEEEauthorblockN{Xiaohui Sun, Jiangwei Mo, Hanlin Wu* \orcidlink{0000-0002-3505-0521}, Jie Ma}
\IEEEauthorblockA{
    \textit{School of Information Science and Technology}\\
    \textit{Beijing Foreign Studies University}\\
    Beijing, China\\
    \{202420229002, 202220119004, hlwu, majie\_sist\}@bfsu.edu.cn
}
}
\maketitle
\begin{abstract}
Recent advancements in diffusion models (DMs) have greatly advanced remote sensing image super-resolution (RSISR). However, their iterative sampling processes often result in slow inference speeds, limiting their application in real-time tasks. To address this challenge, we propose the latent consistency model for super-resolution (LCMSR), a novel single-step diffusion approach designed to enhance both efficiency and visual quality in RSISR tasks. Our proposal is structured into two distinct stages. In the first stage, we pretrain a residual autoencoder to encode the differential information between high-resolution (HR) and low-resolution (LR) images, transitioning the diffusion process into a latent space to reduce computational costs. The second stage focuses on consistency diffusion learning, which aims to learn the distribution of residual encodings in the latent space, conditioned on LR images. The consistency constraint enforces that predictions at any two timesteps along the reverse diffusion trajectory remain consistent, enabling direct mapping from noise to data. As a result, the proposed LCMSR reduces the iterative steps of traditional diffusion models from 50-1000 or more to just a single step, significantly improving efficiency. Experimental results demonstrate that LCMSR effectively balances efficiency and performance, achieving inference times comparable to non-diffusion models while maintaining high-quality output.
\end{abstract}

\begin{IEEEkeywords}
Remote sensing, super-resolution, latent consistency model, single-step diffusion.
\end{IEEEkeywords}

\section{Introduction}
Remote sensing image super-resolution (RSISR) reconstructs low-resolution (LR) images obtained under the limitations of sensor imaging technology and atmospheric conditions~\cite{tu2024rgtgan} to acquire high-resolution (HR) images, thus enabling a wide range of applications such as urban planning~\cite{mao2023elevation}, land cover classification~\cite{xie2021super}, disaster monitoring~\cite{zhao2024see}. Recently, many researchers have utilized deep learning architecture~\cite{wang2022comprehensive}, including the residual dense connections, attention mechanism, to enhance super-resolution (SR) performance. However, the visual quality of the results from these regression-based models is still unsatisfactory.

With further research deepening, generative adversarial networks (GANs)~\cite{goodfellow2014generative} have been proposed to generate visually realistic HR images~\cite{meng2023single,wang2023msagan} and have occupied an important position in the field of SR. However, the GAN training process is complicated and prone to problems such as pattern collapse and unstable training. Recently, diffusion models (DMs), especially the latent diffusion models (LDMs)~\cite{rombach2022high}, which put the diffusion process into latent space with the aim of reducing computation, have shown excellent performance in the field of SR and achieved remarkable results~\cite{wu2023conditional,xiao2023ediffsr}. In SR tasks, DMs generate HR images based on reconstruction of the input LR images through an iterative reverse sampling process, where the samples are progressively denoised during the reconstruction process.

However, a critical limitation of DM-based methods lies in their iterative sampling process, which results in slow inference speeds and hinders real-time applications. To address this issue, researchers have actively explored potential solutions. Song et al.~\cite{song2023consistency} introduced the consistency models, which learn consistency mappings to maintain point consistency along ordinary differential equation (ODE) trajectories. Building on this, Luo et al.~\cite{luo2023latent} integrated the consistency models with LDMs, further advancing the efficiency of image generation. Inspired by these advancements, we propose the latent consistency model for super-resolution (LCMSR) to address the inefficiency of diffusion sampling in SR tasks. 


Specifically, inspired by LDMs~\cite{rombach2022high}, our framework contains two stages. In the first stage, we train a residual autoencoder to capture the differential information between HR and LR images. The encoder maps this differential information into a latent space, generating compact residual representations. The decoder then takes both the latent code and the LR image as inputs, incorporating a dedicated SR branch to restore high-frequency details by integrating the decoded residual information. The second stage focuses on consistency diffusion learning. Here, we generate latent codes to predict residual information, conditioned on the LR image, enforcing that any noisy point on the same ODE trajectory is mapped to the same starting point under consistency constraints, enabling single-step diffusion.

The main contributions can be summarized as follows:

\begin{figure*}[t]
	\centering
	\includegraphics[width=\linewidth]{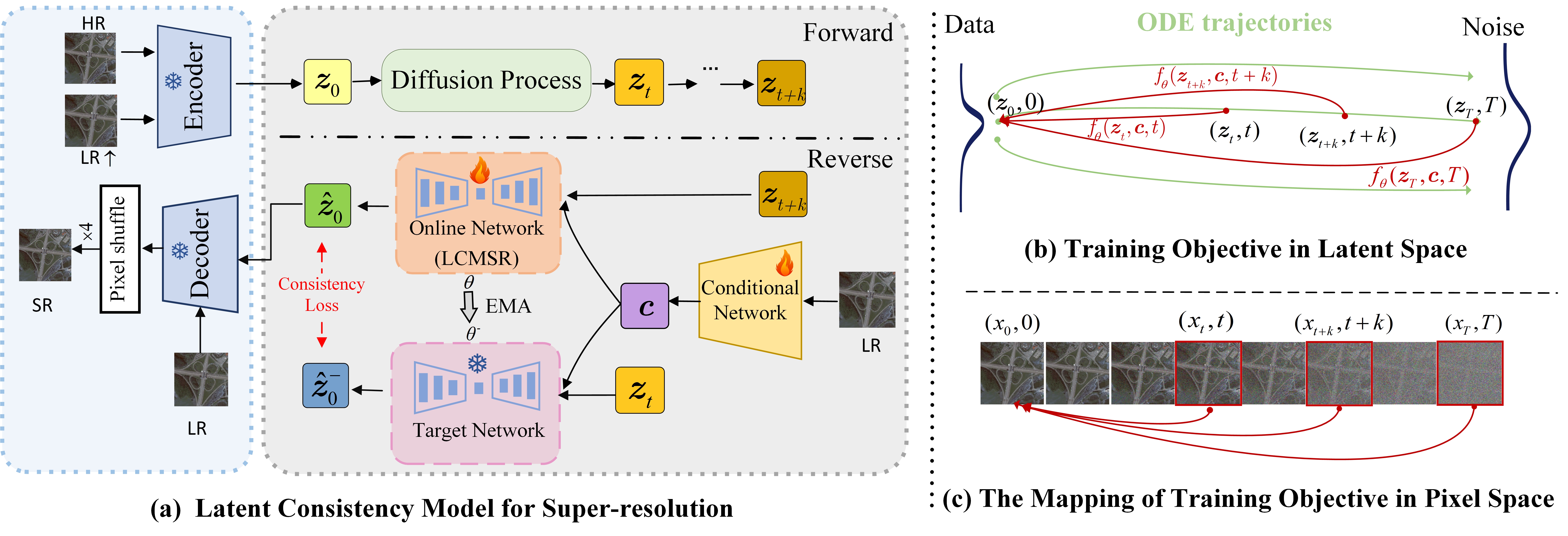}
        \vspace{-1.5em}
	\caption{(a) An overview of our LCMSR framework. LCMSR consists of two parts. The first stage (left) is the pretraining in the latent space, and the second part (right) is the training of the consistency diffusion model. (b) The training Objective is to learn a consistency function $f_{\theta}$, which projects any points on the ODE trajectory to its solution in latent space. (c) Mapping the training objective to the pixel space leads to a more intuitive understanding.}\label{fig:IMRAD}
\end{figure*}

1) We propose LCMSR, first method to apply consistency constraints in latent space, specifically designed for the RSISR task. It reduces the time-step length of DMs from dozens or even thousands to just one step, while maintaining sample quality comparable to state-of-the-art (SOTA) SR models. 
 
2) We employ a residual autoencoder for the RSISR task. This autoencoder exclusively encodes the differential information between HR and LR images, which not only reduces the difficulty of pretraining but also makes it easier for the consistency model to predict latent codes.

\section{Methodology}
The framework of our proposed LCMSR, as illustrated in Fig.\,\ref{fig:IMRAD}(a), achieves single-step diffusion for the SR task by incorporating consistency constraints in the latent space. The proposed model consists of two stages. In the first stage, we pretrain a residual autoencoder to obtain compact representations of the HR-LR differential information. In the second stage, we use a consistency model to learn the data distribution in the latent space.



\subsection{Residual Autoencoder}
\label{sec:autoencoder}

Unlike previous LDMs that directly encode the entire image, our residual autoencoder exclusively encodes the differential information between HR and LR images, which represents the high-frequency details required for SR. Through an encoder-decoder architecture, we obtain a more compact low-dimensional residual encoding of this differential information.

\subsubsection{Encoder}
\label{sec:encoder}
Since LR images contain abundant low-frequency information, we encode only the differences between HR and LR images to reduce the burden of the pretraining stage. For the encoder, the inputs are the HR image and its upsampled LR counterpart, and the output is the residual representation $\bm z$. The architecture of the encoder is identical to that of the LDM \cite{rombach2022high} encoder, except that the number of input channels is doubled. The operation of encoder can be summarized as:
\begin{equation}
\label{eq:1}
\bm z = \mathcal{E} (I_\mathrm{HR}, I_\mathrm{LR}^{\uparrow}),
\end{equation}
where $I_\mathrm{HR}$ and $I_\mathrm{LR}^\uparrow$ represent the HR image and the bicubic upsampled LR image, respectively. 

\subsubsection{Decoder} 
\begin{figure}[b]
	\centering
	\includegraphics[width=\linewidth]{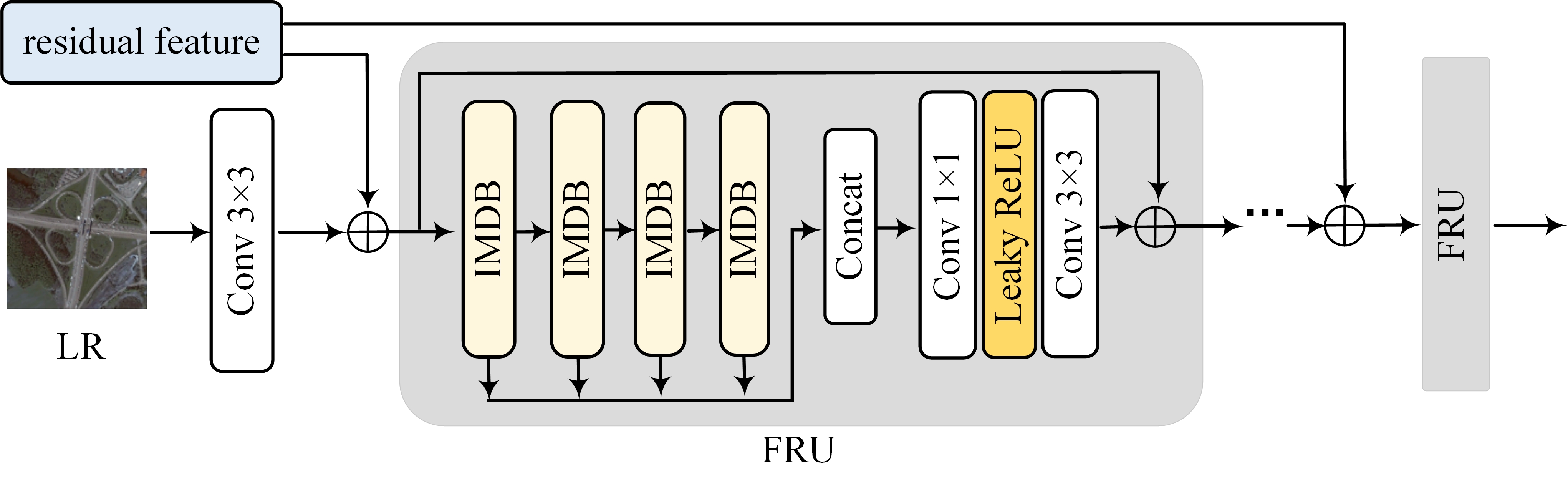}
        \vspace{-1em}
	\caption{Architecture of the SR branch.}\label{fig:SRbranch}
\end{figure}

The decoder is composed of two branches. The first branch is latent representation decoding branch that is symmetric to the encoder, which decodes the generated latent code into the pixel-space. The second branch is the SR branch, whose structure is inspired by \cite{wu2023lightweight}, as illustrated in Fig.\,\ref{fig:SRbranch}.

In the SR branch, the LR image first passes through a $3\times3$ convolutional layer, which serves as a shallow feature extractor. Subsequently, it passes through four sequentially connected feature refinement units (FRUs). Each FRU comprises 12 consecutive lightweight feature-extraction blocks, specifically designed for SR tasks, known as information multi-distillation blocks (IMDBs) \cite{hui2019lightweight}. Additionally, a $1\times 1$ convolutional layer is employed in each FRU to compress the multi-level feature maps. Specifically, each FRU receives side outputs from the latent representation decoding branch as additional information. The additional residual feature maps are directly added to the original feature maps, enhancing the representation of high-frequency details.
Finally, a pixel-shuffle operation is performed to perform a $\times 4$ upsampling. The operation of decoder can be summarized as:
\begin{equation}
    I_\mathrm{SR} = \mathcal{D}(I_\mathrm{LR}, \bm z)
    \label{eq2},
\end{equation}
where $I_\mathrm{SR}$ denotes the final SR result.

\subsection{Latent Consistency Diffusion}
In the latent space, we transform the distribution of the latent representations into a Gaussian distribution through a diffusion process. For the reverse process, we replace the traditional iterative sampling with a consistency model, enabling single-step inference. 



\subsubsection{Forward Diffusion Process} 
The latent representation $\bm z_0 := \bm z$ is obtained using Eq.\,\eqref{eq:1}. In the forward diffusion process, noise is gradually added to the data, and two noisy latent variables $\bm z_t$ and $\bm z_{t + k}$ are obtained in two time steps with a fixed interval $k$:
\begin{align}
    \bm z_{t} &= \sqrt{\bar\alpha_{t}} \bm z_{0} + \sqrt{1-\bar{\alpha}_{t}} \bm \epsilon , \quad \bm \epsilon \sim \mathcal{N}(0, 1)\\
    \bm z _ { t + k} &= \sqrt{ \bar\alpha_{t+k}} \bm z_{0} + \sqrt { 1 - \overline { \alpha } _ { t + k} } \bm \epsilon , \quad \bm \epsilon \sim \mathcal{N}(0, 1),
\end{align}
where $\alpha_{t}=1-\beta_{t}$, $\bar{\alpha}_{t} = \prod_{i=0}^{t} \alpha_{i}$, and $ \left\{\beta_{t}\right\}_{t=1}^{T}$ is a predefined variance schedule.

\subsubsection{Conditional Mechanism} 
In the SR task, the reverse process accept the LR image as a conditioning input. Following previous studies~\cite{wu2023conditional,xiao2023ediffsr}, we extract LR image features as conditional inputs through a dedicated conditional network, thereby accelerating model convergence. The architecture of the conditional network is the same as the encoder in Sec.\ref{sec:autoencoder}. The conditional features $\bm c = \mathrm{CondNet}(I_{\text{LR}}^{\uparrow})$. To better align the conditional features $\bm c$ and the ground truth latent representation $\bm z_{0}$, we define a knowledge distillation (KD) loss with an $L_1$ constraint to accelerate the convergence of the consistency model:
\begin{equation}
\label{eq:6}
 \mathcal{L}_{\mathrm{KD}}(\theta) = \|\mathrm{CondNet}(I_\mathrm{LR}^{\uparrow})-\bm z_{0}\|_1.
\end{equation}

\subsubsection{Consistency Model} 
A key property of the consistency model is that points on the same trajectory converge to the same starting point, as illustrated in Fig.\,\ref{fig:IMRAD}(b) and Fig.\,\ref{fig:IMRAD}(c). The consistency property in the latent space can be expressed as:
\begin{equation}
f(\bm z_{t}, \bm c, t) = f(\bm z_{t+k}, \bm c, t+k), \quad \forall t, t + k \in \left[0, T \right]\label{eq7},
\end{equation}
where $\bm c$ denotes the conditional features extracted from the LR image, $T$ is the total time steps of the diffusion process, and $k$ is a fixed time interval.

To achieve single-step prediction, we aim for the consistency model to map noisy latents at any time step to the starting point, i.e., $ f_{\theta}(\bm z_{t} , \bm c , t) \equiv \bm z_0, \forall t\in [0, T]$. Therefore, we introduce the boundary condition:
\begin{equation}
f_{\theta}(\bm z_{0}, \bm c , 0) = \bm z_0.     
\end{equation}
To satisfy the boundary condition, we parameterize the latent consistency model as follows:
\begin{equation}
f_{\theta}(\bm z_{t}, \bm c,t) = c_{\mathrm{\bm skip}}(t) \bm z_{t} + c_\mathrm{\bm out} (t) \Big( \frac {\bm z_{t} - \sqrt{1-\bar{\alpha}_t}{\bm\epsilon} _ {\theta} ( \bm z_{t} , \bm c , t ) } {\sqrt{\bar{\alpha}_t}} \Big)
\label{eq8},
\end{equation}
where $c_{\mathrm{\bm skip}}(0) = 1, c_\mathrm{\bm out} (0) = 0$, and $\bm{\epsilon} _ { \theta} (\bm z_{t},\bm c,t)$ is a noise predictor of the diffusion model.

Following \cite{song2023consistency}, we stabilize training by updating the target network $\theta^{-}$ using the exponential moving average (EMA) of the trainable parameters from the online denoising network $\theta$. Input $\bm z_{t + k}$ and $\bm z_t$ with varying noise levels to the online network $\bm\epsilon_{\theta}$ and the target network $\bm\epsilon_{\bm\theta^{-}}$ to obtain $\hat{\bm z}_{0}$ and $ \hat{\bm z}_{0}^{-}$, respectively. Next, perform the parameterization by Eq.\,\eqref{eq8}. To ensure the consistency property in Eq.\,\eqref{eq7}, we define the consistency training (CT) loss as follows:
\begin{equation}
    \mathcal{L}_{\mathrm{CT}}(\theta, \theta^{-}) = \|f_{\theta}(\bm z_{t+k}, \bm c, t+k) - f_{\theta^{-}}(\bm z_{t}, \bm c, t)\|_1.
    \label{eq9}
\end{equation}
The training procedure of LCMSR is summarized in Algorithm \ref{algo:LCMSR}.

\begin{algorithm}[t]
    \label{algo:LCMSR}
    \DontPrintSemicolon
    \SetKwInput{KwIn}{Input}
    \SetKwInput{KwOut}{Output}
    \SetKwRepeat{Repeat}{repeat}{until}
    \caption{Latent Consistency Training}
    \KwIn{Dataset $\mathcal{D}=\{(I_\mathrm{HR}, I_\mathrm{LR})\}$, initial model parameter $\theta$, learning rate $\eta$, EMA rate $\mu$, timestep interval $k$, pretrained encoder $\mathcal{E}(\cdot,\cdot)$.}
    \KwOut{Trained model parameters $\theta$.}
    
    Initialize the parameters of $\mathrm{CondNet}(\cdot)$ and $\theta^{-} \leftarrow \theta.$\;
    
    \Repeat{converged}{
        Compute the residual representation:
        \begin{quote} $\bm z = \mathcal{E}(I_{\mathrm{HR}}, I_{\mathrm{LR}}^\uparrow).$\end{quote}
        Sample $\bm\epsilon \sim \mathcal{N}(0, I)$ and $t \sim \mathcal{U}[0, T-k].$\\
        Compute the condition input: $\bm c = \mathrm{CondNet}(I_{\mathrm{LR}}^{\uparrow}).$
        Compute $\mathcal{L}_{\mathrm{CT}}$ and $\mathcal{L}_{\mathrm{KD}}$ by Eq.\,\eqref{eq9} and Eq.\,\eqref{eq:6}.\;
        Update the parameters: 
        $
        \theta \leftarrow \theta - \eta \nabla_{\theta} \mathcal{L}(\theta, \theta^{-}).
        $        
        Update the EMA parameters: 
        \begin{quote}
            $ \theta^{-} \leftarrow \text{stopgrad}(\mu \theta^{-} + (1 - \mu) \theta).$
        \end{quote}
        Take a gradient descent step with
        \begin{quote}
        $\nabla(\mathcal{L}_\mathrm{CT}(\theta, \theta^{-}) + \mathcal{L}_{ \mathrm{KD} }(\theta)).$ 
        \end{quote}
    }
\end{algorithm}




\begin{figure*}[t]
    \centering
    \includegraphics[width=0.95\linewidth]{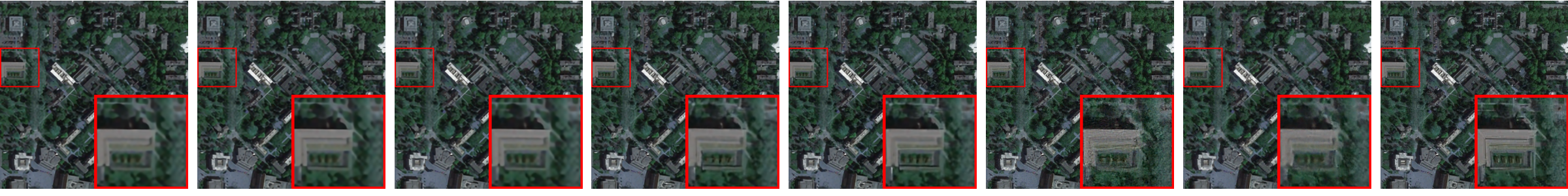}
    \includegraphics[width=0.95\linewidth]{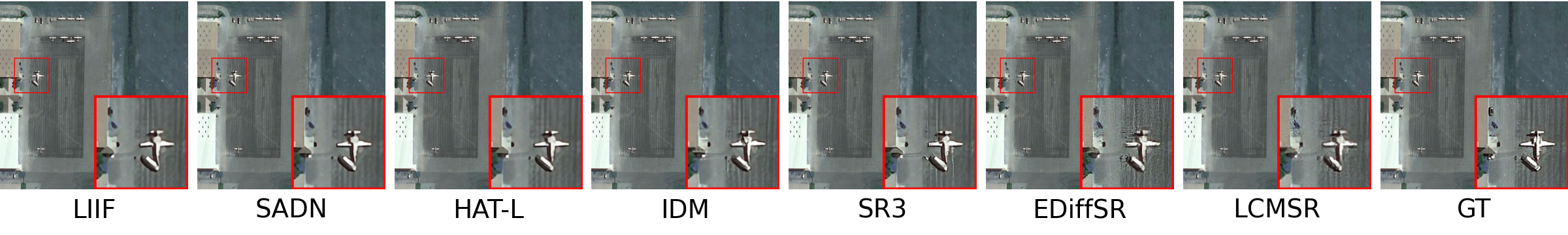}
    \vspace{-1em}
    \caption{Visual comparison of SR results ($\times 4$). The first row is from the AID dataset, and the second row from the DIOR dataset.  Zoomed-in for better detail.}
    \label{fig:visual-comparison}
\end{figure*}

\section{Experimental Results}

\subsection{Datasets}

We evaluate our model on two publicly available datasets: AID~\cite{xia2017aid} and DIOR~\cite{li2020object}. The AID dataset comprises 10,000 HR images spanning 30 categories, split into 9,000 training, 1,000 validation, and 1,000 test images. From DIOR, we randomly select 500 images (800 $\times$ 800) for testing. In all experiments, we employ a scale factor of $\times 4$ and generate LR images using Bicubic interpolation.

\subsection{Implementation Details}

The diffusion process scheduler is configured with 1000 steps, with the variance increasing from $\beta_1 = 0.0015$ to $\beta_{1000} = 0.0155$. The UNet denoising model has 64 base channels. The autoencoder is pretrained for 200 epochs with a mini-batch of 8 and a learning rate of $3.6 \times 10^{-5}$. The first 50 epochs use only $L_1$ loss,  after which the training combines $L_1$, adversarial loss, and regularization loss with weights of $1.0\times10^{-6}$ and $0.5$, respectively. The consistency model is trained from scratch for 200 epochs with a mini-batch of $16$ and an initial learning rate of $8 \times 10^{-5}$. 

\subsection{Comparison with the SOTAs}
We compare our proposed method with several SOTA SR methods, including LIIF \cite{chen2021learning}, SADN \cite{wu2023learning}, HAT-L \cite{chen2023activating}, IDM \cite{gao2023implicit}, SR3 \cite{saharia2022image}, EDiffSR \cite{xiao2023ediffsr}. For a fair comparison, all methods are retrained on the AID training set. We evaluate the performance using three metrics: peak signal-to-noise ratio (PSNR), Fréchet inception distance (FID) \cite{heusel2017gans}, and learned perceptual image patch similarity (LPIPS) \cite{zhang2018unreasonable}.

\begin{table}[t]
  \centering
  \setlength{\tabcolsep}{3.5pt}
  \renewcommand{\arraystretch}{0.9} 
  \caption{Quantitative Comparison of PSNR ($\mathrm{dB}$), FID, and LPIPS on Two Datasets. ``\# Diff. Steps'' Denotes the Number of Diffusion Steps, While ``/'' Denotes Non-Diffusion Models. Best Results Are Shown in \textbf{Bold}, and Second-Best in \textcolor{blue}{Blue}.}
    \begin{tabular}{cccccc}
    \Xhline{1pt}
    \multicolumn{1}{c}{{Datasets}} & \multicolumn{1}{c}{{Methods}} & \multicolumn{1}{c}{{\#\,Diff.\,Steps}} & \multicolumn{1}{c}{{PSNR(dB)\,$\uparrow$}} & \multicolumn{1}{c}{{FID\,$\downarrow$}} & {LPIPS $\downarrow$}\,\bigstrut\\
    \hline
    \hline
    \multirow{8}[2]{*}{AID} & Bicubic &    /   & 27.430 & 59.633 & 0.494 \bigstrut[t]\\
          & LIIF \cite{chen2021learning}  & /     & 29.323 & 42.490 & 0.334 \\
          & SADN \cite{wu2023learning} & /     & \textbf{29.583} & 39.838 & 0.319 \\
          & HAT-L \cite{chen2023activating} & /     & \textcolor{blue}{29.490} & 39.818 & 0.321 \\
          & IDM \cite{gao2023implicit}  & 40    & 27.611 & 29.248 & 0.277 \\
          & SR3 \cite{saharia2022image}   & 40    & 28.150 & 26.014 & \textcolor{blue}{0.252} \\
          & EDiffSR \cite{xiao2023ediffsr} & 100   & 25.238 & \textcolor{blue}{25.968} & 0.273 \\
          & LCMSR (Ours)  & 1     & 27.813 & \textbf{23.393} & \textbf{0.202} \bigstrut[b]\\
    \hline
    \multirow{8}[2]{*}{DIOR} & Bicubic & /     & 26.948 & 65.415 & 0.499 \bigstrut[t]\\
          & LIIF \cite{chen2021learning}  & /     & 29.011 & 34.375 & 0.349 \\
          & SADN  \cite{wu2023learning} & /     & \textbf{29.247} & 32.611 & 0.336 \\
          & HAT-L \cite{chen2023activating} & /     & \textcolor{blue}{29.179} & 32.679 & 0.337 \\
          & IDM \cite{gao2023implicit}  & 40    & 27.557 & 26.985 & 0.284 \\
          & SR3 \cite{saharia2022image}  & 40    & 27.674 & \textcolor{blue}{25.550} & \textcolor{blue}{0.268} \\
          & EDiffSR \cite{xiao2023ediffsr} & 100   & 24.923 & 32.461 & 0.332 \\
          & LCMSR (Ours) & 1     & 27.460 & \textbf{24.112} & \textbf{0.220} \bigstrut[b]\\
    \Xhline{1pt}
    \vspace{-2em}
    \end{tabular}%
  \label{tab:addlabel}%
\end{table}%

Our proposed LCMSR demonstrates strong performance across all datasets while requiring only single diffusion step, significantly fewer than other diffusion-based approaches. Although LCMSR achieves slightly lower PSNR values compared to regression-based methods, it consistently delivers superior perceptual quality, as evidenced by the best FID and LPIPS scores. Fig.\,\ref{fig:visual-comparison} presents a visual comparison of SR results, where LCMSR avoids over-smoothing and generates natural-looking reconstructions that align closely with human perception.

Table~\ref{tab:runtime} demonstrates the efficiency of our proposed consistency model, which reduces the inference time of diffusion-based SR methods to a level comparable with non-diffusion models, achieving a balance between efficiency and performance. 



\begin{table}[tbp]
    \centering
    \caption{Comparison of Average Inference Time and Number of Parameters for Various Methods.}
    \label{tab:runtime}
    \setlength{\tabcolsep}{2.5pt}
    \begin{tabular}{lccccccc}
        \Xhline{1pt}
        & \multicolumn{3}{c}{Regression-based} & \multicolumn{4}{c}{Diffusion-based} \bigstrut[t]\\
        \cmidrule(lr){2-4} \cmidrule(lr){5-8} 
        & LIIF & SADN & HAT-L & SR3 & EDiffSR & IDM & LCMSR \bigstrut[b]\\
        \hline
        \hline
        Runtime (s) & 0.0575 & 0.0581 & 0.2474 & 2.8150 & 13.5648 & 3.0883 & 0.0708 \bigstrut[t]\\
        \# Param. (M) & 22.30 & 7.60 & 40.32 & 92.56 & 26.79 & 111.34 & 31.25 \bigstrut\\
        \Xhline{1pt}
    \end{tabular}%
\end{table}%
 
\begin{table}[t]
    \centering
    \caption{Discussion on Loss Functions on the AID-Tiny Dataset Consisting of 15 Images. All Models Are Trained For 100 Epochs for Fast Evaluation. Best Results Are Shown In \textbf{Bold}.}
    \setlength{\tabcolsep}{11pt}{
        \begin{tabular}{lccc}
            \Xhline{1pt}
            Models & PSNR(dB) $\uparrow$ & FID $\downarrow$   & LPIPS  $\downarrow$ \bigstrut\\
            \hline
            \hline
            w/o KD loss & 28.320 & 80.513 & 0.225  \bigstrut[t]\\
            w/o consistency loss & \textbf{29.310} & 101.636 & 0.271 \\
            Ours  & 28.450 & \textbf{77.170} & \textbf{0.202} \bigstrut[b]\\
            \Xhline{1pt}
        \end{tabular}%
    }
    \label{tab:ablation}%
\end{table}%

\subsection{Ablation Study}
We investigate the impact of loss functions on model performance. First, we explore an alternative to single-step diffusion without consistency loss, where an $L_1$ constraint is applied between the model output $f_{\theta}(\bm z_t, \bm c, t)$ and the ground truth $\bm z_0$. Second, we investigate the contribution of the KD loss. As shown in Table \ref{tab:ablation}, the consistency loss ensures high visual quality, as evidenced by lower FID and LPIPS metrics, while the KD loss is essential for accelerating convergence, making both components indispensable for our proposal.

\section{Conclusion}
In this work, we propose LCMSR, a latent consistency model specifically designed for RSISR. By introducing consistency constraints into the diffusion process, LCMSR enables rapid inference through a direct mapping from any noisy latent code to the ODE trajectory’s starting point. To better adapt to the characteristics of remote sensing images, we train the consistency model from scratch, eliminating the dependency on pretrained diffusion models. This approach significantly reduces the inference steps from thousands to just one. Experimental results demonstrate that LCMSR achieves an excellent balance between efficiency and performance.

\small
\bibliographystyle{IEEEtranN}
\bibliography{references}

\end{document}